\documentclass[11pt,twoside]{article}  % Leave intact
\usepackage{asp2006}
\usepackage{adassconf}

\begin{document}   % Leave intact

%-----------------------------------------------------------------------
%			    Paper ID Code
%-----------------------------------------------------------------------
% Enter the proper paper identification code.  The ID code for your paper 
% is the session number associated with your presentation as published 
% in the official conference proceedings.  You can find this number by 
% locating your abstract in the printed proceedings that you received 
% at the meeting, or on-line at the conference web site.
%
% This identifier will not appear in your paper; however, it allows different
% papers in the proceedings to cross-reference each other.  Note that
% you should only have one \paperID, and it should not include a
% trailing period.

\paperID{F4}

%-----------------------------------------------------------------------
%		            Paper Title 
%-----------------------------------------------------------------------
% Enter the title of the paper.

\title{Youpi, a Web-based Astronomical Image Processing Pipeline}
       
%-----------------------------------------------------------------------
%          Short Title & Author list for page headers
%-----------------------------------------------------------------------
% Please supply the author list and the title (abbreviated if necessary) as 
% arguments to \markboth.
%
% The author last names for the page header must appear in one of 
% these formats:
%
% EXAMPLES:
%     LASTNAME
%     LASTNAME1 and LASTNAME2
%     LASTNAME1, LASTNAME2, and LASTNAME3
%     LASTNAME et al.
%
% Use the "et al." form in the case of four or more authors.
%
% If the title is too long to fit in the header, shorten it: 
%
% EXAMPLE: change
%    Rapid Development for Distributed Computing, with Implications for the Virtual Observatory
% to:
%    Rapid Development for Distributed Computing

\markboth{Monnerville and S\'emah}{Youpi, a Web-based Astronomical Image Processing Pipeline}

%-----------------------------------------------------------------------
%		          Authors of Paper
%-----------------------------------------------------------------------
% Enter the authors followed by their affiliations.  The \author and
% \affil commands may appear multiple times as necessary.  List each
% author by giving the first name or initials first followed by the
% last name. Do not include street addresses and postal codes, but 
% do include the country name or abbreviation. 
%
% If the list of authors is lengthy and there are several institutional 
% affiliations, you can save space by using the \altaffilmark and \altaffiltext 
% commands in place of the \affil command.

\author{M.\ Monnerville \& G.\ S\'emah}
\affil{Terapix, Institut d'Astrophysique de Paris, CNRS, University of Pierre et Marie Curie, UMR 7095, Paris, F-75014, France}

%-----------------------------------------------------------------------
%			 Contact Information
%-----------------------------------------------------------------------
% This information will not appear in the paper but will be used by
% the editors in case you need to be contacted concerning your
% submission.  Enter your name as the contact along with your email
% address.

\contact{Mathias Monnerville}
\email{monnerville@iap.fr}

%-----------------------------------------------------------------------
%		      Author Index Specification
%-----------------------------------------------------------------------
% Specify how each author name should appear in the author index.  The 
% \paindex{ } should be used to indicate the primary author, and the
% \aindex for all other co-authors.  You MUST use the following syntax: 
%
%    \aindex{LASTNAME, F.~M.}
% 
% where F is the first initial and M is the second initial (if used). Please 
% ensure that there are no extraneous spaces anywhere within the command 
% argument. This guarantees that authors that appear in multiple papers
% will appear only once in the author index. Authors must be listed in the order
% of the \paindex and \aindex commmands.

%\paindex{Djorgovski, S.}
%\aindex{King, I.~R.}
%\aindex{Biemesderfer, C.~D.}

\paindex{Monnerville, M.}
\aindex{S\'emah, G.}

%-----------------------------------------------------------------------
%			Subject Index keywords
%-----------------------------------------------------------------------
% Enter up to 6 keywords that are relevant to the topic of your paper.  These 
% will NOT be printed as part of your paper; however, they will guide the creation 
% of the subject index for the proceedings.  Please use entries from the
% standard list where possible, which can be found in the index for the 
% ADASS XVI proceedings. Separate topics from sub-topics with an exclamation 
% point (!). 


% We reset the footnote counter for the hyperlink since it does not
% appear to recognize the previous 3 footnotes generated from the
% altaffilmarks.  

\setcounter{footnote}{3}

%-----------------------------------------------------------------------
%			       Abstract
%-----------------------------------------------------------------------
% Type abstract in the space below.  Consult the User Guide and Latex
% Information file for a list of supported macros (e.g. for typesetting 
% special symbols). Do not leave a blank line between \begin{abstract} 
% and the start of your text.

\begin{abstract} 
Youpi\footnote{http://youpi.terapix.fr/} stands for ``YOUpi is your processing PIpeline''. 
It is a portable, easy to use web application providing high level functionalities to perform 
data reduction on scientific FITS images. It is built on top of open source processing tools 
that are released to the community by Terapix, in order to organize your data on a computer 
cluster, to manage your processing jobs in real time and to facilitate teamwork by allowing 
fine-grain sharing of results and data. On the server side, Youpi is written in the \emph{Python}
programming language and uses the \emph{Django}\footnote{http://www.djangoproject.com/} web 
framework. On the client side, Ajax techniques are used along with the \emph{Prototype} and 
\emph{script.aculo.us} Javascript librairies\footnote{http://www.prototypejs.org/, http://script.aculo.us/}.
\end{abstract}

%-----------------------------------------------------------------------
%			      Main Body
%-----------------------------------------------------------------------
% Place the text for the main body of the paper here.  You should use
% the \section command to label the various sections; use of
% \subsection is optional.  Significant words in section titles should
% be capitalized.  Sections and subsections will be numbered
% automatically. 

\section{Introduction}

Youpi is a new generic and versatile web-based pipeline that is suitable to handle both large 
surveys, like the CFHTLS (Goranova et al 2009), and smaller, less-well organized sets of 
observations. 

Youpi is designed to run and supervise data management by keeping track of all data locations 
and parameters involved during the processing life cycle. It starts by ingesting FITS image
information into the database and manages the processing all the way to the end products (image stacks,
catalogues).

\section{Handling FITS Data Ingestion}

Ingesting images is the main entry point for Youpi users. Thanks to a dedicated selection widget,
the user chooses one or several data paths holding FITS images (available either locally or on
the network  by using network shares such as NFS), which prompts Youpi to read the headers and ingest 
the relevant information in the database. Youpi reports back to the user by e-mail as soon as
ingestion is completed. Ingestion can be carried out once for all, so that every ingested image
is available for processing at any time. Youpi provides user-friendly tools to query images and data
(see section \ref{sec:ims}). 

\section{Data Processing}

Youpi is a high-level tool acting as a wrapper for freely available, low-level software packages:
it currently supports the \emph{QualityFITS} image quality assessment software, the \emph{SCAMP}
astrometric and photometric calibration tool (Bertin 2006), the \emph{SWarp} image resampling
and stacking software (Bertin et al. 2002) and the \emph{SExtractor} source extraction and catalogue production tool
(Bertin\&Arnouts 1996).

All these processings modules are internally available as plug-ins. Some of them may be disabled 
to perform any specific or fragmented processing sequence requested by the user. The 
plug-in architecture is modular and versatile. A detailed guide on how to write custom plug-ins and
extend processing capabilities is available on the Youpi website.

Youpi manages data access rights and protections by implementing a Unix-like file permission system. 
Each user has a personal password-protected account, where data sets and results can have 
user-defined permissions for the owner, a group of owners and the rest of the world. Permissions
can be changed to share images, data products or even pipeline configuration files.

\section{Data Management}

The various widgets and tools of the Youpi user interface have been painstakingly designed to
simplify the selection and organisation of data, as well as the management and set up of
configuration and result files.

\paragraph{The Data Path Selector}

The path selector may be invoked from any processing plug-in to select and save data paths
in real time (e.g. to provide a path to external {\tt .ahead} files for calibrating a bunch of
images using the \emph{SCAMP} plug-in). The path selector widget can browse the user's local
-- and remote, if network mounts are used -- filesystems, and save those directories for later
reuse.

\paragraph{The Image Selector}\label{sec:ims}

\begin{figure}[h]
\epsscale{1}
\plotone{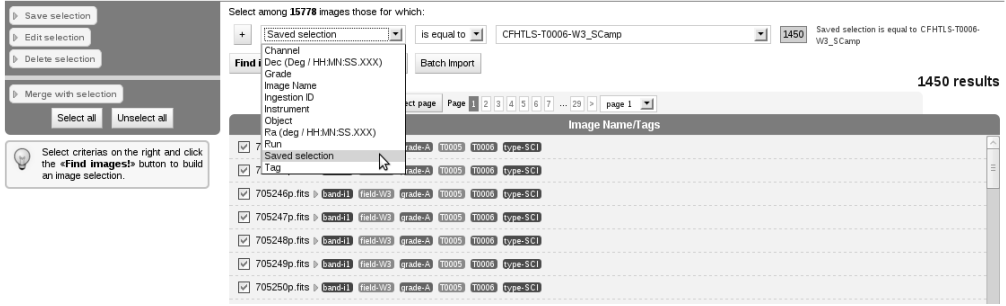}
\caption{A selection of 1450 images is built by retrieving all images from the 
\emph{CFHTLS-T0006-W3\_Scamp} saved selection.}\label{fig:ims}
\end{figure}

The image selector (Fig. \ref{fig:ims}) sets up image lists based on previously 
ingested images. Multiple search criteria can be combined together, including observing run IDs,
filters/channels, grades (Youpi provides an integrated interface for image quality assessment), \ldots

Once images are selected, basic editing commands can be used to save the current 
selection. Saved selections can also be deleted or merged with others. In order to increase the 
image selector's flexibility, options have been added to upload a plain text file describing the 
selection (file names and optional checksums) or to batch-import all text files within 
a directory. These two features allow Youpi to create and save selections in a semi-automated way,
rather than manually. This comes in particularly handy when processing surveys with a very large
number of images as all image selections may be prepared beforehand.

\paragraph{Organizing Data With Tags}

\begin{figure}[h]
\epsscale{1}
\plotone{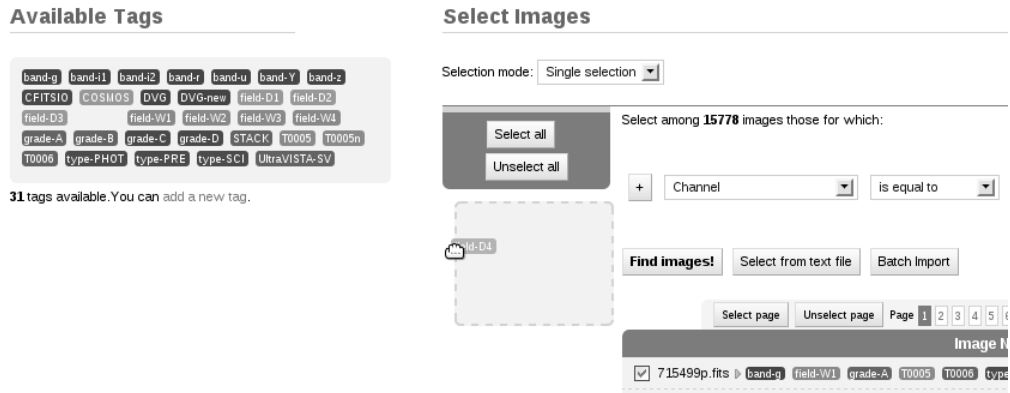}
\caption{Dragging the \emph{field-D4} tag over a selection of images.}\label{fig:tags}
\end{figure}

A tag is a non-hierarchical keyword that can be assigned to a piece of information. Youpi allows 
tagging ingested images in order to add some kind of metadata useful for describing items. Tagged 
images can later be searched by tag name using the image selector. Applying tags on a selection of 
images is straightforward (Fig. \ref{fig:tags}): all it takes is to create a new tag -- or use
an existing one -- and drag-and-drop it over a dedicated drop zone to mark or unmark the current
selection. The number of tags which may be applied to an image is practically unlimited.

\paragraph{The Processing Cart}

Before running any processing with Youpi, all input data, configuration parameters and processing 
tasks are first bundled together in one processing cart item. The processing cart acts like shopping 
carts on commercial websites. This is a convenient mechanism for sharing processings among users -- 
processing items can be saved for later use and loaded back to the processing cart at 
anytime by others -- or for setting run-time options before submitting jobs on the cluster.

\section{Cluster Computing With Condor}

Youpi relies on the \emph{Condor}\footnote{http://www.cs.wisc.edu/condor/} software framework
for distributing jobs on the local machine, a cluster of computers, or a compute grid.
Ingestions as well as processing cart items are submitted as Condor jobs. Youpi handles all 
input parameters and generates a \emph{Condor submission file} -- a plain text file specifying 
job requirements, environments and complete command lines -- ready to be executed on the system.

Depending on the requirements, it may be useful to target a specific subset of cluster 
nodes. Youpi has a page dedicated to Condor requirements setup, from which custom policies or
selections may be defined. Custom policies are \emph{dynamic rules}: several criteria can be 
defined using regular expressions; they will only be computed just before the job is
submitted. In contrast, custom selections are made of \emph{static nodes} 
selected among the available cluster nodes displayed on screen.

Jobs submitted to the Condor cluster can be monitored directly from your browser on the
Youpi \emph{Active Monitoring} page. Important information such as the job 
description, remote host, running time, owner and current status are displayed and refreshed 
in real time, without reloading the page.

\section{Summary}

Youpi is a versatile pipeline which can be used on a single host -- a desktop computer or even 
a laptop -- or on larger installations such as computer clusters. The software is actively maintained
at Terapix\footnote{http://terapix.iap.fr/} and published under an 
open source license. Youpi was first released to the general community on September 2009.
The current version offers
native support for data from the CFHT-MegaCam, CFHT-WIRCam and ESO-VISTA-VIRCAM instruments,
and we plan to support more data sources in the future. 

\acknowledgments

We are grateful to E.\ Bertin, Y.\ Goranova, P.\ Hudelot, F.\ Magnard, H.\ McCracken, Y.\ Mellier and 
M.\ Schultheis for active support and testing. We would also like to thank Y.\ Kakazu for her Japanese
translation of the Youpi abstract.

% Do not place any material after the references section

\end{document}